# Failed theories of superconductivity

Jörg Schmalian

Department of Physics and Astronomy, and Ames Laboratory, Iowa State University, Ames, IA 50011, USA

**Almost half a century passed between the discovery of superconductivity by Kamerlingh Onnes and the theoretical explanation of the phenomenon by Bardeen, Cooper and Schrieffer. During the intervening years the brightest minds in theoretical physics tried and failed to develop a microscopic understanding of the effect. A summary of some of those unsuccessful attempts to understand superconductivity not only demonstrates the extraordinary achievement made by formulating the BCS theory, but also illustrates that mistakes are a natural and healthy part of the scientific discourse, and that inapplicable, even incorrect theories can turn out to be interesting and inspiring.**

The microscopic theory of superconductivity was developed by Bardeen, Cooper and Schrieffer (BCS)[1]. It was published in 1957, 46 years after the original discovery of the phenomenon by Kamerlingh Onnes[2]. During those intervening years numerous scientists tried and failed to develop an understanding of superconductivity, among them the brightest minds in theoretical physics. Before 1957 other correct descriptions of the phenomenon superconductivity were developed. Those include the work by Cornelius Gorter and Hendrik Casimir in 1934[3] and most notably the phenomenological theories of Heinz and Fritz London in 1935[4] and Vitaly Ginzburg and Lev Landau in 1950[5]. The triumph of the BCS-theory was, however, that it gave an explanation that started from the basic interactions of electrons and atoms in solids, i.e. it was a microscopic theory, and, at the same time, explained essentially all of the complex properties of conventional superconductors. For a number of observables, the agreement between theory and experiment is of unparalleled precision in the area of interacting many body systems.

When discussing failed attempts to understand superconductivity, we must keep in mind that they are a natural and healthy part of the scientific discourse. They are an important part of the process of finding the right answers. These notes are not written to taunt those who tried and did not succeed. On the contrary, it is the greatness that comes with names like Joseph John Thompson, Albert Einstein, Niels Bohr, Léon Brillouin, Ralph Kronig, Felix Bloch, Lev Landau, Werner Heisenberg, Max Born, and Richard Feynman that demonstrates the dimension of the endeavor undertaken by John Bardeen, Leon N Cooper and J. Robert Schrieffer. Formulating the theory of superconductivity was one of the hardest problems in physics of the $20^{th}$ century.

In light of the topic of this article, it is not without a sense of irony that the original discovery by Kamerlingh Onnes seems to have been motivated, at least in part, by an incorrect theory itself, proposed by another highly influential thinker. Lord Kelvin had argued, using the law of corresponding states, that the resistivity of all substances diverges at sufficiently low temperatures[6]. Kamerlingh Onnes was well aware of Kelvin's work, and while he might have been skeptical about it, the proposal underscored the importance to investigate the $T \rightarrow 0$ limit of the electric resistivity. For a discussion of Kelvin's role in motivating Kamerlingh Onnes' experiment, see Ref[7].

We know now that superconductivity is a macroscopic quantum effect. Even though many elements of the new quantum theory were developed by Planck, Bohr, Sommerfeld, Einstein and others starting in 1900, it is clear in hindsight that it was hopeless to explain superconductivity before the formulation of quantum mechanics by Heisenberg[8] and Schrödinger[9] during 1925-1926. This was crisply articulated by Albert Einstein when he stated during the 40th anniversary of Kamerlingh Onnes's professorship in Leiden in 1922: *"with our far-reaching ignorance of the quantum mechanics of composite systems we are very far from being able to compose a theory out of these vague ideas"*[10]. The vague ideas resulted out of his own efforts to understand superconductivity using the concept of *"molecular conduction chains"* that carry supercurrents. What Einstein had in mind resembles to some extend the soliton motion that does indeed occur in one dimensional conductors[11,12]. In his view *"supercurrents are carried through closed*

*molecular chains where electrons undergo continuous cyclic exchanges"*. Even though no further justification for the existence of such conduction paths was given, the approach was based on the view that superconductivity is deeply rooted in the specific chemistry of a given material and based on the existence of a state that connects the outer electrons of an atom or molecule with those of its neighbors. Einstein also suggested an experiment to falsify his theory: bringing two different superconducting materials in contact, no supercurrent should go from one superconductor to the other, as the molecular conducting chains would be interrupted at the interface. Again, it was Kamerlingh Onnes who performed the key experiment and showed that supercurrents pass through the interface between lead and tin, demonstrating that Einstein's theory was incorrect, as was stated in a *post scriptum* of Einstein's original paper[10]. For historical accounts of Einstein's work on superconductivity and his impact on condensed matter physics in general, see Ref[13] and Ref[14], respectively.

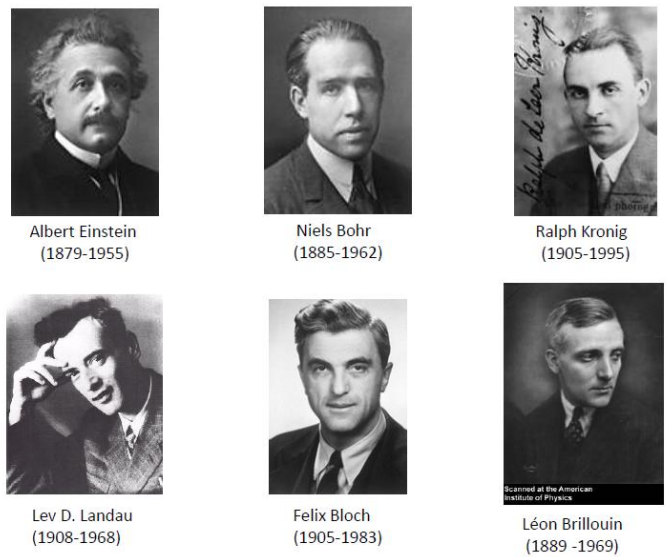

**Figure 1: Einstein, Bohr, Kronig, Landau, Bloch, and Brillouin.made proposals for microscopic theories of superconductivity prior to the ground breaking experiment by Meissner and Ochsenfeld in 1934.**

While Einstein's concept of molecular conduction chains did not turn out to be the right one, he was correct in insisting that a theory of superconductivity cannot be based on the concept of non-interacting electrons in solids. It is also remarkable to read the introductory paragraph of his paper on superconductivity, where he states that *"...nature is a merciless and harsh judge of the theorist's work. In judging a theory, it never rules 'Yes' in best case says 'Maybe', but mostly 'No'. In the end, every theory will see a 'No'."* Just like Einstein, most authors of failed theories of superconductivity were well aware that their proposals were sketchy and their insights of preliminary character at best. Einstein's theory should not be considered as a singular intuition, on the spur of the moment. Joseph John Thompson, who discovered the electron in his famous cathode ray experiment[15], had already made a proposal to explain superconductivity in terms of fluctuating electric dipole chains in 1915[16]. In 1921, Kamerlingh Onnes proposed a model of superconducting filaments[17]. Below the superconducting transition temperature, conduction electrons would *"slide, by a sort of association, through the metallic lattice without hitting the atoms"*. In judging these early ideas about superconductivity, one must appreciate that even the normal state transport properties of metals were only poorly understood. In fact the bulk of Einstein's paper is concerned with a discussion of the normal state electric and heat conductivities.

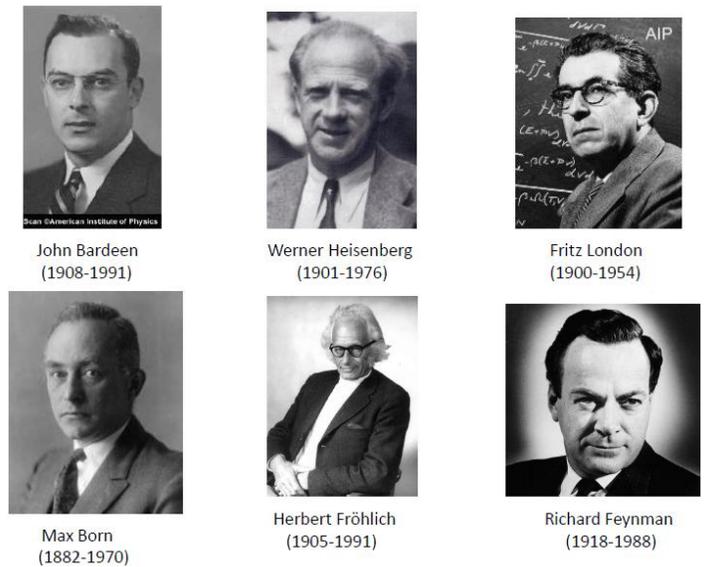

**Figure 2: Between the second world war and the formulation of the BCS theory, unsuccessful attempts to formulate microscopic theories of superconductivity were made by Bardeen, Heisenberg, London, Born, Fröhlich, and Feynman.**

After the formulation of quantum mechanics, motivated to a large extent by the properties of single electrons and atomic spectra, it soon became clear that this new theory explained phenomena that were much more complex. Heisenberg's important contributions to the theory of magnetism[18] and Felix Bloch's lasting work on the theory of electrons in crystals[19] were already published during 1928. Soon after, an understanding of ordinary electrical conductors and of the peculiar thermal properties of solids was developed using the new quantum theory (see Ref.[20] for an historical account). The main tools needed to formulate the theory of superconductivity were now or would soon become available. Given the swift success in other areas of solid state physics, the inability to formulate a theory of superconductivity demonstrated the need for conceptually new insights that were required to solve this problem.

Of the "failures" to explain superconductivity during the early post quantum mechanics days, two theories are noteworthy for their elegance and the distinguished participants involved. Those are the spontaneous current approach independently proposed by Felix Bloch and Lev

Landau[21] and the theory of coherent electron-lattice motion by Niels Bohr and, around the same time, by Ralph de Laer Kronig[22,23]. It may be said that the authors of these ideas worked on wrong theories, but certainly not that they proposed uninteresting ideas. Both concepts had some lasting impact or relevance.

In the Drude formula for electric transport[24] the conductivity is given as

$$\sigma = \frac{e^2 n}{m}\tau, \tag{1}$$

with electron charge $e$, mass $m$, density $n$ and scattering time $\tau$, respectively. This result led to the frequently proposed view that superconductors are perfect conductors, i.e. $\sigma \to \infty$, due to a vanishing scattering rate $\tau^{-1}$. In his 1933 paper Landau gave a very compelling argument that superconductors should not be considered as perfect conductors[21]. His reasoning was based upon the fact that the resistivity right above $T_c$ is still finite, i.e. $\tau$ is finite and electrons must be undergoing scattering events. Landau stressed that a mechanism based on the notion of a perfect conductor requires that $\tau^{-1}$ jumps discontinuously from a finite value to zero. He argued that it is highly implausible that all interactions are suddenly switched off at the transition temperature. The remarkable aspect of this argument is that it was almost certainly made without knowledge of the crucial experiment by Meissner and Ochsenfeld[25] that ruled against the notion of superconductors as perfect conductors. In addition, Landau's formulation of the theory in 1933, albeit wrong, contained the first seeds of the eventually correct Ginzburg-Landau theory of superconductivity[5]. Having rejected the idea of superconductivity due to an infinite conductivity, he analyzed the possibility of equilibrium states with finite current. Landau proposed to expand the free energy of the system in powers of the current **j**[21]:

$$F(\mathbf{j}) = F(\mathbf{j}=\mathbf{0}) + \tfrac{a}{2}\mathbf{j}^2 + \tfrac{b}{4}\mathbf{j}^4. \tag{2}$$

No odd terms occurred in $F(\mathbf{j})$ as the energy should not depend on the current direction. The equilibrium current $\langle\mathbf{j}\rangle$ is given by the value that minimizes $F(\mathbf{j})$. Landau argued that $b>0$, to ensure a continuous transition, and that the coefficient of the quadratic term changes sign at the transition temperature $a \propto T-T_c$, to allow for a finite equilibrium current, $\langle\mathbf{j}\rangle \neq \mathbf{0}$, below $T_c$. He pointed out that the resulting heat capacity jump agrees with experiment, but cautioned that the temperature variation of the current, $|\mathbf{j}| \propto (T_c-T)^{1/2}$, seems to disagree with observations. The approach was inspired by the theory of ferromagnetism and already used the much more general concept of an order parameter that discriminates between different states of matter. Landau's first order parameter expansion of the free energy of antiferromagnets was published in 1933[26], after his manuscript on superconductivity. The widely known Landau theory of phase transition was only published in 1937[27]. It is amazing that these lasting developments seem to have been inspired by an incorrect theory of superconductivity. Landau expansions have since been successfully used to describe the critical point of water, liquid crystals, the properties of complex magnets, the inflationary cosmic evolution right after the big bang, and many other systems.

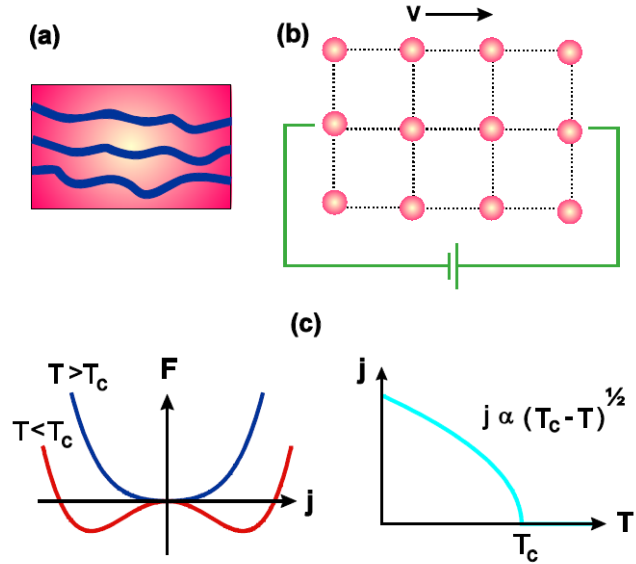

**Figure 3: (a) Sketch of Einstein's molecular conduction chains. (b) Kronig's electron crystal that was supposed to slide as a whole in an external electric field. (c) Landau's expansion of the free energy with respect to the equilibrium current.**

Felix Bloch did not publish his ideas about coupled spontaneous currents, which were, just like Landau's theory, motivated by the theory of ferromagnetism. Yet, through his efforts he became highly knowledgeable about the status of the experimental observations in the field. During the early 1930's Bloch's infamous, somewhat cynical, second theorem, that every theory of superconductivity can be disproved, was frequently cited among theorists[20]. It reflected the degree of despair that must have existed in the community. His first theorem, which was, in contrast to the second one, perfectly serious, was concerned with the energy of current carrying states. The theorem contained the proof that Bloch's own theory of coupled spontaneous currents was in fact false. A summary of Bloch's first theorem was given by David Bohm in 1949[28]: Suppose a finite momentum $\langle \mathbf{P} \rangle = \langle \Psi | \mathbf{P} | \Psi \rangle \neq \mathbf{0}$ in the ground state $\Psi$ of a purely electronic system, which leads to a finite current $\langle \mathbf{j} \rangle = e \langle \mathbf{P} \rangle / m$. Let the Hamiltonian

$$H = \sum_i \left( -\frac{\hbar^2 \nabla^2}{2m} + U(\mathbf{r}_i) \right) + \sum_{i<j} V(\mathbf{r}_i - \mathbf{r}_j) \tag{3}$$

consist of the kinetic energy, the potential $U(\mathbf{r}_i)$ due to the ion lattice and the electron-electron interaction $V(\mathbf{r}_i - \mathbf{r}_j)$. One then finds that the wave function $\Phi = \exp(i \delta \mathbf{p} \cdot \sum_i \mathbf{r}_i / \hbar) \Psi$ has a lower energy than $\Psi$ if the variational parameter $\delta \mathbf{p}$ points opposite to $\langle \mathbf{P} \rangle$. Thus, $\Psi$ cannot be the ground state unless $\langle \mathbf{j} \rangle = \mathbf{0}$ for the purely electronic problem, Eq.(3). This result clearly invalidated Landau's and Bloch's proposals.

The second idea proposed in 1932 by Bohr and Kronig was that superconductivity would result from the coherent quantum motion of a lattice of electrons. Given Bloch's stature in the field, theorists like Niels Bohr where eager to discuss their own ideas with him. In fact Bohr, whose theory for superconductivity was already accepted for publication in the July 1932 issue of the journal "Die Naturwissenschaften", withdrew his article in the proof stage, because of Bloch's criticism (see Ref.[20]). Kronig was most likely also aware of Bloch's opinion when he published his ideas[22]. Only months after the first publication he responded to the criticism made by Bohr and Bloch in a second manuscript[23]. It is tempting to speculate that his decision to publish and later defend his theory was influenced by an earlier experience: in 1925 Kronig proposed that the electron carries spin, i.e. possesses an internal angular momentum. Wolfgang Pauli's response to this idea was that it was interesting but incorrect, which discouraged Kronig from publishing it. The proposal for the electron spin was made shortly thereafter by Samuel Goudsmit and George Uhlenbeck[29]. Kronig might have concluded that it is not always wise to follow the advice of an established and respected expert.

In his theory of superconductivity Kronig considered the regime where the kinetic energy of electrons is sufficiently small such that they crystallize to minimize their Coulomb energy. Given the small electron mass he estimated that the vibrational frequencies of this electron crystal were much higher than those of crystal lattice vibrations, amounting to its strong rigidity. In the presence of an electric field, this electron crystal was then supposed to slide as a whole, as the rigidity of the electron crystal would suppress scattering by individual electrons. The superconducting transition temperature would correspond to the melting point of the electron crystal. Furthermore, the suppression of superconductivity with magnetic field could, in Kronig's view, be explained due to the interference of field induced circular orbits with the crystalline state. Clearly, the constructed approach was a sophisticated version of the idea of a perfect conductor. The flaw of the approach was that it overestimated quantum zero-point fluctuations that were supposed to prevent the crystal from getting pinned. Still, it is noteworthy that the proposal used the rigidity of a macroscopic state - the electron crystal - to avoid single electron scattering. As Kronig pointed out, the notion of an electron lattice and its potential importance for electron transport and superconductivity were already voiced by Frederick Lindemann in 1915[30] and refined by J. J. Thompson in 1922[31]. It is nevertheless remarkable that Kronig used the concept of an electron crystal within a quantum mechanical approach two years prior to Eugene Wigner's pioneering work on the subject[32].

Léon Brillouin, who made key contributions to quantum mechanics, solid state physics, and information theory, proposed his own theory of superconductivity during the spring of 1933[33]. He assumed an electronic band structure $\varepsilon(\mathbf{p})$ with a local maximum at some intermediate momentum $\mathbf{p}_0$. $\mathbf{p}_0$ is neither close to $\mathbf{p} = \mathbf{0}$ nor, to use the contemporary terminology, at the

Brillouin zone boundary. He showed that the equilibrium current of such a system vanishes, but that non-equilibrium populations $n(\mathbf{p})$ of momentum states, may give rise to a net current. He then argued that due to the local maximum in $\varepsilon(\mathbf{p})$ such non-equilibrium states have to overcome a barrier to relax towards equilibrium, leading to metastable currents. At higher temperatures, equilibration becomes possible, causing those supercurrents to relax to zero. Brillouin realized that his scenario naturally implied a critical current. Since Brillouin argued that superconductivity was a metastable state, his proposal was at least not in conflict with Bloch's first theorem. In 1934, Gorter and Casimir gave strong evidence for the fact that superconductivity is an equilibrium phenomenon[3], which ruled out Brillouin's approach.

Before discussing further examples of "failures", the breakthrough experiment by Walter Meissner and Robert Ochsenfeld[25], followed by the pioneering theory of Heinz and Fritz London[4], must be mentioned. Meissner and Ochsenfeld demonstrated that the magnetic flux is expelled from a superconductor, regardless of its state in the distant past. The phenomenological theory that naturally accounted for this finding was soon after proposed by the London brothers in their 1934 paper in the Proceedings of the Royal Society[4]. To appreciate some of the later proposals for microscopic theories of superconductivity, we deviate from the theme of this manuscript and briefly summarize this correct and lasting phenomenological theory: London and London discussed that in a perfect conductor, Ohm's law, $\mathbf{j} = \sigma \mathbf{E}$, is replaced by the acceleration equation

$$\frac{\partial \mathbf{j}}{\partial t} = \frac{ne^2}{m} \mathbf{E}, \tag{4}$$

appropriate for the frictionless motion of a charged particle in an electric field $\mathbf{E}$. Combining this relation with Faraday's law yields

$$\nabla \times \frac{\partial \mathbf{j}}{\partial t} = \frac{ne^2}{mc} \frac{\partial \mathbf{B}}{\partial t}, \tag{5}$$

with magnetic induction $\mathbf{B}$. From Ampere's law follows then that the magnetic field decays at a length scale $\lambda_L$ to its initial value $\mathbf{B}_0$ where $\lambda_L^{-2} = 4\pi ne^2/(mc^2)$. London and London concluded that in case of a perfect conductor *"the field $\mathbf{B}_0$ is to be regarded as 'frozen in' and represents a permanent memory of the field which existed when the metal was last cooled below the transition temperature."* If correct, cooling from the normal state to the superconducting state in an external field would keep the magnetic induction at its finite high temperature value $B_0 \neq 0$, directly contradicting the Meissner-Ochsenfeld experiment. In addition, it would imply that superconductors are not in thermodynamic equilibrium as their state depends on the system's history in the distant past. These facts led London and London to abandon the idea of a perfect conductor. They further realized that the history dependence of the field value is a direct consequence of the presence of time derivatives on both sides of Eq.(5), while the desired decay of the field on the length scale $\lambda_L$ follows from the curl operator on the left hand side of Eq.(5). Instead of using Eq.(5), they simply proposed to drop the time derivatives, leading to a new material's equation for superconductors:

$$\nabla \times \mathbf{j} = \frac{ne^2}{mc}\mathbf{B}.\qquad(6)$$

This phenomenological London equation guarantees that the field decays to zero on the length $\lambda_L$, referred to as the London penetration depth. The authors concluded that *"in contrast to the customary conception that in a supraconductor a current may persist without being maintained by an electric or magnetic field, the current is characterized as a kind of diamagnetic volume current, the existence of which is necessarily dependent upon the presence of a magnetic field."* In other words, London and London recognized that superconductors are perfect diamagnets. With the Nazi regime coming into power in 1933 a sudden shift of the research efforts from Germany to the United States and England took place, changing the priorities of numerous researchers. While the theory of the London brothers is a beautiful phenomenological account of the Meissner effect, a microscopic theory was not immediately inspired by this key experiment.

After the Second World War, Werner Heisenberg, one of the creators of modern quantum mechanics, took serious interest in formulating a theory of superconductivity[34]. His theory was based on the assumption that strong Coulomb interactions dramatically alter the character of electrons. Instead of forming plane waves, electrons near the Fermi energy would localize. The model was treated using a variational single electron wave function. Heisenberg realized that a crucial challenge for any theory of superconductivity was to derive the tiny observed energy advantage of the superconducting state, compared to other possible states. His search for new bound states in the vicinity of the Fermi energy was quite original and clearly pointed into the right direction. His confidence regarding his own results is nevertheless impressive: *"If ... condensation takes place through the Coulomb forces, one can scarcely think of any other mechanism which reduces ordinary Coulomb energies to so small values."*[35]. The shortcoming in the approach was that Heisenberg did not accept that the Meissner effect is at odds with the infinite conductivity approach to superconductivity. He stated: *"the essential difference from several more recent attempts is the assumption that the perfect conductivity rather than the diamagnetism is the primary feature of the phenomenon"*[35]. Heisenberg claimed to have derived the London equation from his theory. Following his calculation one finds that the derivation is based on the implicit assumption that the initial magnetic field value vanishes, i.e. $\mathbf{B}_0 = \mathbf{0}$.

Max Born, in collaboration with Kai Chia Cheng, proposed a theory of superconductivity in 1948[36]. One goal of the theory was to explain why some metals are superconducting while others are not. The authors gave empirical evidence for the fact that superconductors tend to have a Fermi surface located in the near vicinity of the Brillouin zone boundary, which suggested to them that ionic forces, i.e. those due to $U(\mathbf{r}_i)$ in Eq.(3) would play an important role. In their analysis, which is essentially a density functional theory, they found that the electron-electron interaction causes changes in the occupation of momentum states near the Brillouin zone boundary. Apparently unaware of the fact that Bloch's first theorem is valid even in the presence of the ionic potential, they claimed, without giving further details, to have found asymmetric occupations with a net momentum and therefore a spontaneous current. Born and Cheng were correct to point out that interactions affect the distribution in momentum space, and become important for systems with large portions of the Fermi surface close to the Brillouin zone boundary. Yet, a correct analysis of their own formalism should have led them to the conclusion that asymmetric occupations do not correspond to the lowest energy state.

Fritz London strongly disagreed with Heisenberg and Born. Being challenged by those ideas he formulated his own microscopic theory of superconductivity[37]. In an 1948 paper he first demonstrated that Heisenberg's theory cannot yield superconductivity as a matter of principle. He then pointed out that Heisenberg missed the leading term in his treatment of the Coulomb interaction, which, in London's view, is the exchange interaction, discovered by Heisenberg himself many years before[18]. London then proposed that this exchange interaction, when it favors ferromagnetism, is responsible for superconductivity as it can lead to an *"attraction in momentum space"*. In materials known then, superconductivity is not caused by this effect, but rather by vibrations of the crystal lattice, i.e. sound excitations. On the other hand, the superfluid state of ³He[38-40] and superconductivity in a number of strongly correlated electron systems have been discussed in terms of an exchange of collective degrees of freedoms, such as ferromagnetic, antiferromagnetic or current fluctuations[40-44], that bear some resemblance to what London had in mind. Of course, our understanding of these materials is based on the BCS theory. London's ideas, while interesting, remained vague. The most remarkable aspect of this 1948 paper by Fritz London was however that it emphasized clearly that a superconductor is a macroscopic object in a coherent quantum state: *"the superconductor as a pure quantum mechanism of macroscopic scale"*. This view was also one of the key elements of the elegant phenomenological theory by Ginzburg and Landau that was proposed soon thereafter [5].

If we talk about the creators of unsuccessful theories of superconductivity we should not ignore that Bardeen himself belonged, at least temporarily, to this honorable club. One attempt to solve the problem, briefly mentioned in an abstract in Physical Review, goes back to 1941 and is based on the idea that electrons couple to small periodic distortions of the lattice[45]. In his scenario, the electron-lattice interaction would cause small lattice distortions with large unit cells containing approximately $10^6$ atoms, leading to $10^6$ bands in a tiny Brillouin zone. Bardeen estimated that approximately $10^4$ of those bands were located in the vicinity of $\pm k_B T_c$ of the Fermi energy. Since some bands would have a very small mass, he expected them to contribute to a large diamagnetic response. He further estimated that a fraction of approximately $10^{-6}$ of the electrons would participate in this diamagnetic shielding. As the reasoning is based upon lattice distortions, Bardeen concluded that the electron-lattice coupling of good superconductors should be large, a conclusion that also followed from the BCS theory, yet for very different reasons.

Another key experiment that elucidated the origin of superconductivity was the measurement of the isotope effect at Rutgers University in the group of Bernard Serin[46] and by Emanuel Maxwell[47] at the National Bureau of Standards, Washington, D. C. in 1950. These experiments found a change in the transition temperature upon changing the ion mass via isotope substitution, and demonstrated that vibrations of the crystalline lattice were closely tied to the emergence of superconductivity. The isotope effect was in fact predicted by Herbert Fröhlich theoretically[48]. In 1950 Bardeen and, independently, Fröhlich worked on the problem and came to the conclusion that vibrations of the crystalline lattice lead to a net attraction between electrons and are a likely cause of superconductivity[48,49]. This attraction is present even if one includes the electron-electron Coulomb interaction as shown by Bardeen and Pines[50]. For an historical account of this period of time, see Ref.[51]. Starting in 1950 one should not refer to Bardeen's or Fröhlich's ideas as failed, rather as incomplete. The link between the established attraction between electrons due to sound excitations and superconductivity was still unresolved. For example, in 1953 Fröhlich developed a beautiful theory for one dimensional system with electron lattice coupling[52] and obtained in his calculation a gap in the electron spectrum, as was in fact seen in a number of experiments. The theory is a perfectly adequate and very interesting starting point for the description of sliding charge density waves. We know now that the one-dimensionality makes the model somewhat special and that the gap is incomplete in three dimensions. Most importantly, the theory did not explain superconductivity and did not account for the Meissner effect. Still, there is strong evidence that sliding charge density waves, as proposed by Fröhlich to explain superconductivity, have in fact been observed in quasi-one dimensional conductors; see for example Refs.[53,54].

Richard Feynman is another scientist who battled with the problem of superconductivity. When he was asked in 1966 about his activities in the 1950's he answered: *"there is a big vacuum at that time, which is my attempt to solve the superconductivity problem -- which I failed to do"*[55,56]. Feynman's views, shortly before the BCS theory appeared, are summarized in the proceedings of a meeting that took place the fall of 1956[57]. He stated that the Fröhlich-Bardeen model was, in his view, the correct approach; it only needed to be solved adequately. Even by performing a very accurate analysis using the method of perturbation theory, i.e. by including the vibrations of the crystal without fundamentally changing the properties of the electrons, Feynman was still unable to obtain a superconducting state. From this calculation he concluded that the solution must be beyond the scope of perturbation theory, which turned out to be correct. The formulation of such a theory was accomplished by John Bardeen, Leon N Cooper and J. Robert Schrieffer.

After one goes over these unsuccessful attempts to understand superconductivity, there seem to be a few natural conclusions that one can draw: first, the development of the BCS theory is among the most outstanding intellectual achievement in theoretical physics. Second, theories emerge quite differently from how they are taught in the classroom or presented in textbooks, and, finally, an inapplicable theory, even a wrong theory, can be interesting, inspiring and may turn out to be useful in another setting.

The author acknowledges Paul C. Canfield, Rafael M. Fernandes, Jani Geyer, and Vladimir G. Kogan for a critical reading of the manuscript.

[1]J. Bardeen, L. N. Cooper and J. R. Schrieffer, Physical Review **106**, 162 (1957).

[2] H. Kamerlingh Onnes, Communications from the Physical Laboratory of the University of Leiden **29**, 1 (1911).

[3] C. S. Gorter and H.Casimir, Zeitschrift für Technische Physik **15**, 539 (1934).

[4]H. London and F. London, Proceedings of the Royal Society A **149**, 71 (1935).

[5]V. L. Ginzburg and L. D. Landau, Soviet Physics JETP **20**, 1064 (1950).

[6] Lord Kelvin, Philosophical Magazine **3**, 257 (1902).

[7] P. H. E. Mejier, American Journal of Physics **62**, 1105 (1994).

[8]W. Heisenberg, Zeitschrift für Physik **33**, 879, (1925), ibid. **39**, 499 (1926).

[9] E. Schrödinger, Annalen der Physik **79**, 361 (1926); ibid. **79**,489 (1926); ibid. **81**, 109 (1926).

[10]A. Einstein "Theoretische Bemerkungen zur Supraleitung der Metalle", in Het Natuurkundig Laboratorium der Rijksuniversiteit te Leiden in de Jaren 1904-1922 (Eduard Ijdo, Leiden) 429 (1922).

[11]W. P. Su, J. R. Schrieffer, and A. J. Heeger, Physsical Review B **22**, 2099 (1980).

[12]A. J. Heeger, S. Kivelson, J. R. Schrieffer and W.-P. Su, Reviews of Modern Physics **60**, 781 (1988).

[13]T. Sauer, Archive for History of Exact Science  **61**, 159 (2007).

[14]M. Cardona, Albert Einstein as the father of solid state physics, in 100 anys d'herència Einsteiniana, pp. 85-115, edited by: P. González-Marhuenda, Universitat de València, (2006) [arXiv:physics/0508237].

[15]J. J. Thomson, Philosophical Magazine **44**, 293 (1897).

[16] J. J. Thompson, Philosophical Magazine **30**, 192 (1915).

[17] H. Kamerlingh Onnes, Communications from the Physical Laboratory of the University of Leiden. Supplement **44a**, 30 (1921).

[18] W. Heisenberg, Zeitschift für Physik **49**, 619 (1928).

[19]F. Bloch, Zeitschift für Physik **52**, 555 (1928).

[20]L. Hoddeson, G. Baym, and M. Eckert, Reviews of Modern Physics **59**, 287 (1987).

[21] L. D. Landau, Physikalische Zeitschrift der Sowjetunion **4**, 43 (1933).

[22]R. Kronig, Zeitschrift für Physik **78**, 744 (1932).

[23]R. Kronig, Zeitschrift für Physik **80**, 203 (1932).

[24]P. Drude, Annalen der Physik **306**, 566 (1900); ibid. **308**, 369 (1900).

[25]W. Meissner and R. Ochsenfeld, Naturwissenschaften **21**,787 (1933).

[26]L. D. Landau, Physikalische Zeitschrift der Sowjetunion **4**, 675 (1933).

[27]L. D. Landau, Soviet Physics JETP **7**, 19 (1937); ibid. **7**, 627 (1937).

[28]D. Bohm, Physical Review **75**, 502 (1949).

[29]S. Goudsmit and G.E. Uhlenbeck, Physica **6**, 273 (1926); G.E. Uhlenbeck and S. Goudsmit, Naturwissenschaften **47**, 953 (1925).

[30] F. A. Lindemann, Philosophical Magazine **29**, 127 (1915).

[31]J. J. Thompson, Philosophical Magazine **44**, 657 (1922).

[32]E. P. Wigner, Physical Review **46**, 1002 (1934).

[33]L. Brillouin, Comptes Rendus Hebdomadaires des Seances de L Academie des Sciences **196**, 1088 (1933); J. Phys. Radium **4**, 333 (1933); ibid. **4**, 677 (1933).

[34]W. Heisenberg, Zeitschrift f. Naturforschung **2a**, 185 (1947); ibid. **3a**, 65 (1948).

[35]The Electron Theory of Superconductivity, in Two Lectures by W. Heisenberg, Cambridge Univ. Press (1949).

[36]M. Born and K. C. Chen, Nature **161**, 968 (1948); ibid. **161**,1017 (1948).

[37]F. London, Physical Review **74**, 562 (1948).

[38] A. J. Leggett, Reviews of Modern Physics **47**, 331 (1975).

[39]J.C. Wheatley, Reviews of Modern Physics **47**, 415 (1975).

[40]D. Vollhardt and P. Wölfle, The Superfluid Phases Of Helium 3, Taylor & Francis (1990).

[41] D.J. Scalapino, Journal of Low Temperature Physics **117**, 179 (1999).

[42] T. Moriya, K. Ueda, Reports on Progress in Physics **66**, 1299 (2003).

[43] A. V. Chubukov, D. Pines, J. Schmalian, Physics of Superconductors 1, ed. by K.H. Bennemann and J.B. Ketterson (Springer), (2003).

[44] Y. Yanasea, T. Jujob, T. Nomurac, H. Ikedad, T. Hotta, and K. Yamada, Physics Reports **387**, 1 (2003).

[45] J. Bardeen, Physical Review **59**, 928 (1941).

[46] C. A. Reynolds, B. Serin, W. H. Wright, and L. B. Nesbitt, Physical Review **78**, 487 (1950).

[47] E. Maxwell, Physical Review **78**, 477 (1950); ibid. **79**, 173 (1950).

[48] H. Fröhlich, Proceedings of the Physical Society **A63**, 778 (1950); Physical Review **79**, 845 (1950); Proceedings of the Physical Society **A64**, 129 (1951).

[49] J. Bardeen, Physical Review **79**, 167 (1950); ibid. **80**, 567 (1950); ibid. **81**, 829 (1951).

[50] J. Bardeen and D. Pines, Physical Review **99**, 1140 (1955).

[51] L. Hoddeson, Journal of Statistical Physics **103**, 625 (2001).

[52] H. Fröhlich, Proceedings of the Royal Society **A223**, 296 (1954).

[53] N. Harrison, C.H. Mielke, J. Singleton, J. Brooks, M. Tokumoto, Journal of Physics: Condensed Matter **13**, L389 (2001).

[54] G. Blumberg, P. Littlewood, A. Gozar, B. S. Dennis, N. Motoyama, H. Eisaki, S. Uchida, Science **297**, 584 (2002).

[55] Interview of Richard F. Feynman, by Charles Weiner, June 28 1966, p. 168, American Institute of Physics, College Park, MD.

[56] D. Goodstein and J. Goodstein, Physics in Perspective **2**, 30 (2000).

[57] R. P. Feynman, Reviews of Modern Physics **29**, 205 (1957).